\documentclass[amssymb,aps,floatfix,floats,preprintnumbers,prb,showpacs,twocolumn]{revtex4-1}
\usepackage{graphicx,hyperref,rcs}
\usepackage{mathptmx,times}
\usepackage[applemac]{inputenc}
\graphicspath{{figures/}}

\RCS $Id: msh.tex,v 1.9 2011/08/02 14:15:17 bs Exp $
\newcommand{\no}{\nonumber}
\newcommand{\be}{\begin{equation}}
\newcommand{\ee}{\end{equation}}

\newcommand{\bea}{\begin{eqnarray}}
\newcommand{\eea}{\end{eqnarray}}
\newcommand{\bd}{\begin{displaymath}}
\newcommand{\ed}{\end{displaymath}}
\newcommand{\ba}{\begin{array}}
\newcommand{\ea}{\end{array}}
\newcommand{\bi}{\begin{itemize}}
\newcommand{\ei}{\end{itemize}}
\newcommand{\bc}{\begin{center}}
\newcommand{\ec}{\end{center}}
\newcommand{\bfl}{\begin{flushleft}}
\newcommand{\efl}{\end{flushleft}}
\newcommand{\bfr}{\begin{flushright}}
\newcommand{\efr}{\end{flushright}}

\newcommand{\vQ}{{\bf Q}}
\newcommand{\vk}{{\bf k}}

\newcommand{\ms}{$m_\vQ(h)$} 
\newcommand{\mm}{$m_0(h)$}

\def\C{$\rm{Cu(pz)_2(ClO_4)_2}$}
\def\Cb{$\bf{Cu(pz)_2(ClO_4)_2}$}
\begin{document}

\title{Staggered moment dependence on field-tuned quantum
fluctuations in 2D frustrated antiferromagnets }

\author {Mohammad Siahatgar, Burkhard Schmidt, and  Peter Thalmeier }

\affiliation{ 
Max Planck Institute for the  Chemical Physics of Solids, D-01187
Dresden, Germany
}

\begin{abstract}
We propose an efficient method to identify the degree of frustration
in quasi-2D antiferromagnets described by the $J_1-J_2$ Heisenberg
model.  The frustration ratio $J_2/J_1$ is usually obtained from
analysis of susceptibility, specific heat and saturation field.  We
show that the non-monotonic field dependence of the staggered moment
caused by the suppression of quantum fluctuations in a field depends
strongly on the frustration ratio.  This gives a powerful criterion to
determine $J_2/J_1$ using a combination of exact diagonalization (ED)
method for finite clusters and spin wave analysis.  We apply this
method to the quasi-2D compound \C{} and show that it leads to an
intermediate ratio $J_2/J_1\simeq 0.2$ for the frustration.  We also
explain the observed anomalous increase of transition temperature in
applied fields as an effect of reduced quantum fluctuations.
\end{abstract}


\preprint{\RCSId}

\maketitle
%%%%%%%%%%%%%%%%%%%%%%%%%%%%%%%%%%%%%%%%%%%%%%%%%%%%%%%%%%%%%%%%%%%%%%%%%%
%

\section{Introduction}
\label{sect:introduction}

Quasi-two dimensional antiferromagnets may show a subtle interplay
between quantum fluctuations and exchange frustration.  This is
clearly apparent in the 2D square lattice $J_1-J_2$ model which has
been found to be approximately realized in a class of layered $V^{2+}$
($S=\frac{1}{2}$) compounds \cite{melzi:00,kini:06}.  An anisotropic
version of this model is also relevant for Fe pnictides
\cite{schmidt:10}.  The ground state of the model is determined by the
size of the frustration angle $\phi=\tan^{-1}(J_2/J_1)$  ($-\pi\leq\phi\leq\pi$).
Depending on $\phi$, N\'eel as well as
columnar antiferromagnetic (NAF/CAF) structures with wave vectors \vQ
=($\pi,\pi$) or ($\pi,0$) respectively may be realized.  The size of
staggered moment in these phases is reduced from the classical value
1/2 by the effect of quantum fluctuations in the ground state.  It
depends strongly on the frustration angle and vanishes around
$\phi/\pi\simeq 0.15$ ($J_2/J_1=0.5$) where a nonmagnetic stacked
dimer state \cite{singh:99} and close to $\phi/\pi\simeq 0.85$
($J_2/J_1=-0.5$) where a spin nematic ground state \cite{shannon:06}
appears.  Therefore the determination of $\phi$ is of central
importance to characterize frustrated 2D quantum magnets.  Frequently
it is obtained by comparing results of high temperature series
expansion or finite temperature Lanczos method (FTLM) to the
temperature dependence of susceptibility and magnetic specific heat
\cite{misguich:03,shannon:04}.  Analysis of the saturation fields is
also employed \cite{tsirlin:09b}.  However, in contrast to the
determination of a single $J$ for the nonfrustrated model, the former
method is inaccurate and even ambiguous \cite{misguich:03,shannon:04}
for the frustrated case. It is not able to discriminate between
frustration angles $\phi$ and $\frac{\pi}{2}-\phi$, in particular this means
if J$_2$/J$_1$$<0$ the cases $J_1>0,J_2<0$ and  $J_1<0,J_2>0$ cannot be distinguished.

Here we propose a powerful method for the determination of the
frustration ratio in 2D quantum antiferromagnets in each sector.  Quantum
fluctuations depend on the canting angle between the moments, it was
shown in Ref.~\cite{thalmeier:08} that this leads to a nonlinear
uniform magnetization characteristic for the degree of frustration.
An even more drastic effect may occur in the staggered moment
\cite{luscher:09}.  The classical canting together with field
suppressed quantum fluctuation lead to a non-monotonic field
dependence which depends crucially on $\phi$.  Comparison of
theoretical prediction from numerical ED and spin wave theory with
experimental results of the staggered moment field dependence  $m_\vQ(H)$ 
can precisely determine the frustration ratio in the NAF and CAF sector.  As
an example this method is demonstrated for the 2D Cu-pyrazine compound
\C{}. We show that a fit to  $m_\vQ(H)$ and the
 FTLM fit to  $\chi(T)$ both give values for the frustration angle $\phi$ which are considerably larger
 than reported previously although they are still deep within the NAF region. We demonstrate that
 the former method gives a reliable value for $\phi$ in contrast to the latter.
In addition the anomalous field dependence of the transition temperature $T_N(H)$
is explained within a self-consistent RPA spin wave theory as a result of frustration
suppressed by the field.

\section{Structure factor of the $J_1-J_2$ model}
\label{sect:structure}

\begin{figure}

\centerline{\includegraphics[width=.9\columnwidth,angle=0]{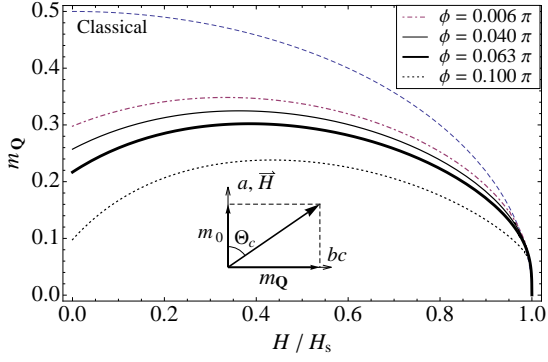}}
\caption{Field dependence of staggered moment from spin wave theory
for several 
frustration angles $\phi$. When $\phi$ approaches the (classical)
critical value $\phi/\pi=0.15$ where
AFM at zero field breaks down, \ms{} behaves strongly non-monotonic.}
\label{fig:Fig1}
\end{figure}

Our analysis is based on the quasi-2D frustrated square lattice spin S=1/2  model
including the Zeeman term: \bea \mathcal{H} &=& J_{1}\sum_{{\langle ij \rangle}}
\vec S_i \cdot \vec S_j + J_{2}\sum_{{\langle \langle ij \rangle \rangle}} \vec
S_i \cdot
\vec S_j \no\\
&&+ J'\sum_{\langle ij \rangle_{\perp}} \vec S_i \cdot \vec S_j -g\mu_B\vec
H\sum_i\vec S_i
\label{eq:ham}
\eea Here $J_1=J_c\cos\phi$ and $J_2=J_c\sin\phi$ are the in-plane nearest and
next nearest neighbor exchange constants with $J_c=(J_1^2+J_2^2)^\frac{1}{2}$
giving the overall energy scale. Furthermore $J'$ is the coupling between 2D
layers with $J'/J_c\ll 1$. The latter term will only be needed for analysis of
3D AF transition temperature.  The field dependent total moment is given by \bea
m_t^2(H)=m_0^2(H)+m_\vQ^2(H) \eea consisting of uniform and staggered moment
perpendicular and parallel to the plane respectively (inset of
Fig.~\ref{fig:Fig1}). Within the numerical ED Lanczos approach they can be
expressed in terms of the static structure factors by $m_0^2=S_{zz}(0)$ and
$m_\vQ^2=\frac{1}{2}\bigl[S_{xx}(\vec Q)+S_{yy}(\vec Q)\bigr]$ using the
definition
 \bea
S_{\alpha\beta}(\vec q,H) =\frac{1}{\mathcal{N}_{\alpha\beta}}
\sum_{i,j=1}^N
\left\langle S_i^\alpha S_j^\beta\right\rangle e^{i\vec q(\vec R_i-\vec R_j)}.
	\label{eq:struct}
\eea The expectation values should be evaluated with the corresponding
ground-state at field $H$. The uniform and staggered moments are obtained by
scaling the structure factor with the size $N$ of the clusters. At $H = 0$, the
spin (S) dependent normalization factor is $\mathcal{N}_{\alpha\beta}= N(N+1/S)$ as explained in  
Ref.~\onlinecite{schmidt:10}.
It must be modified for finite magnetic fields to ${\cal N}_{zz}=N^2$ and ${\cal
N}_{xx}={\cal N}_{yy}=N(N-1)$ to account for the effect of dropping non-zero
on-site terms for the staggered moment in the above sum, Eq.~(\ref{eq:struct}),
which is necessary to achieve the correct limiting values at the saturation
field for arbitrary tile size $N$.
% %%%%%%%%%%%%%%%%%%%%% figure %%%%%%%%%%%%%%%%%%%%%%%%%%%%%%%%%%%
%%%%%%%%%%%%%%%%%%%%%%fig%%%%%%%%%%%%%%%%%%%%%%%%%%%%%%%%%%%%%
%

\section{Uniform and staggered moment from linear spin wave theory and exact diagonalization}
\label{sect:moments}

It is instructive to calculate both moments also within spin wave approximation
for comparison. In the classical limit one has (in units of $g\mu_B$):
$m_0=S\cos\Theta_c$, $m_\vQ=S\sin\Theta_c$ where $\cos\Theta_c=H/H_s$ is the
classical canting angle of spins as counted from the field oriented  $\parallel
a$ or perpendicular to the quasi-2D $bc$ plane . We use this convention to be
compatible with  \C{} discussed later. For the NAF case the saturation field  is
given by $H_s=4J_1/(g\mu_B)$ which is independent of $J_2$. In the classical
picture the staggered moments of constant size $m_t(H)=S$ are simply tilted out
of the plane until at $H_s$ only the uniform moment $m_0=S$ is left
(inset of Fig.~\ref{fig:Fig1}). For H = 0 quantum fluctuations  reduce the size of the
moment  and the amount of reduction strongly depends on the frustration degree
given by $\phi$. However, at saturation ($H=H_s$) the total moment always has to
return to the classical value $m_t = m_0=S$ because the fully polarized
(ferromagnetic) state is an eigenstate and quantum fluctuations are absent. This
means that the field dependence of  \ms, \mm{} and likewise of $m_t(H)$ is tuned
by $\phi$. Here $m_0(h)$ is the {\em uniform\/} moment or magnetization which
was shown in Ref.~\onlinecite{thalmeier:08} to exhibit nonlinear field variation
depending on the size of $\phi$. Similarly the field variation of the {\em
staggered\/} moment $m_\vQ(h)$ should be strongly influenced by the size of
frustration. We propose that this effect may be used as powerful means to determine
$\phi=\tan^{-1}(J_2/J_1)$.
%  %%%%%%%%%%%%%%%%%%%%% figure %%%%%%%%%%%%%%%%%%%%%%%%%%%%%%%%%%%
\begin{figure}

\centerline{\includegraphics[width=0.8\columnwidth,angle=0]{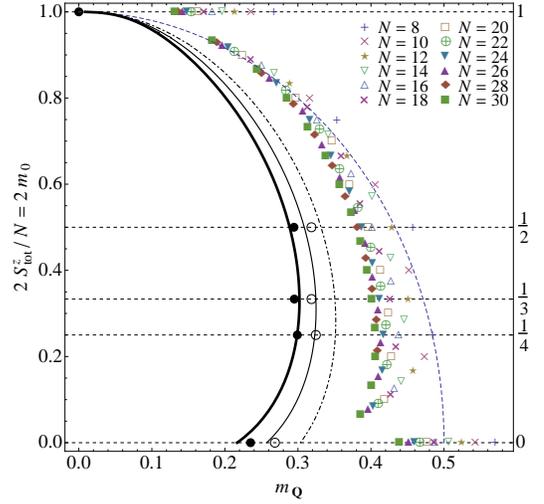}}
\caption{Parametric representation of uniform (\mm) and staggered
(\ms) moment as function of field. Here $H=0$ at the lower right and
$H=H_s$ at the upper left corner. Dashed line: classical result.
Symbols: ED values for finite tiles with $N= 8-30$ and
$\phi/\pi=0.063$. Empty (full) circles: extrapolated values from finite size
scaling for $\phi/\pi=0.04$ ($\phi/\pi=0.063$). Full thin (thick) line: spin wave result for  intermediate frustration
$\phi/\pi=0.04$ ($\phi/\pi=0.063$). Dash-dotted line: spin wave result for small frustration
$\phi/\pi=0.006$.}
\label{fig:Fig2}
\end{figure}
% %%%%%%%%%%%%%%%%%%%%%fig%%%%%%%%%%%%%%%%%%%%%%%%%%%%%%%%%%%%%

The Hamiltonian (\ref{eq:ham}) may be diagonalized in the local coordinate
system of the canted spins~\cite{schmidt:10} leading to spin wave energies 
\bea
\label{eq:sw}
E_\vk(h)&=&\Bigl\{ [A_\vk-B_\vk\cos^2\Theta_c]^2- [B_\vk(1-\cos^2\Theta_c)]^2
\Bigl\}^\frac{1}{2}\no\\
A_\vk&=&J_\vk +\frac{1}{2}\left(J_{\vk+\vQ}+J_{\vk-\vQ}\right)
-2J_{\vQ}\\
B_\vk&=&J_\vk -\frac{1}{2}\left(J_{\vk+\vQ}+J_{\vk-\vQ}\right) \no 
\eea 

where $J_\vk=2J_1(\cos k_y+\cos k_z)+ 4J_2\cos k_y\cos k_z$ is the Fourier transform of the
intra- layer exchange couplings and \vk~ and \vQ~ are wave vectors lying in the bc-plane.
Here we set $J'=0$. Because $J'/J\ll 1$ it is not important for ground state properties. 
Defining $h\equiv g\mu_BH$ we have
$\cos\Theta_c=h/h_s $ with $h_s=2SA_{0}$. The uniform and staggered moments are
then given by the ground-state expectation value of the total moment projected
onto the field direction and the plane perpendicular to the field, respectively,
leading to
\begin{eqnarray}
    m_{0}
    &=&
    \cos\Theta_{\text c}
    \left[
    S+\frac{1}{2}\frac{1}{N}\sum_{\bf k}
    \frac{B_{\bf k}\left(A_{\bf k}-B_{\bf k}\right)}{A_{0}E_{\bf
k}(h)}
    \right],
    \\
    m_{\vQ}
    &=&
    \sin\Theta_{\text c}
    \left\{
    S-
    \frac{1}{2\sin^2\Theta_{\text c}}
    \left[
    \frac{1}{N}\sum_{\bf k}
    \frac{A_{\bf k}-B_{\bf k}\cos^{2}\Theta_{\text c}}{E_{\bf k}(h)}
    -1
    \right]
    \right.
    \nonumber\\
    &\phantom{=}&
    \phantom{\sin\Theta_{\text c}}
    \left.
    {}
    -
    \frac{\cos^2\Theta_{\text c}}{2\sin^2\Theta_{\text c}}
    \frac{1}{N}\sum_{\bf k}
    \frac{B_{\bf k}\left(A_{\bf k}-B_{\bf k}\right)}{A_{0}E_{\bf
k}(h)}
    \right\}
    \label{eq:ms}
\end{eqnarray}
up to ${\cal O}(1/S)$.  The calculated staggered moment for various
frustration angles $\phi$ is shown in Fig.~\ref{fig:Fig1}.
It is obvious from this figure that the field dependence
of the staggered moment \ms{} is strongly influenced by the frustration
angle.  In contrast to the classical case it exhibits non-monotonic
behavior.  This appears because firstly the size of the total moment
increases with field due to suppression of quantum fluctuation and
secondly the moments are canted out of the bc plane, which reduces the
staggered projection.  For larger $\phi$ one approaches the region of
the nonmagnetic phase ($\phi/\pi\simeq 0.15$) where spin wave theory
eventually breaks down ($m_\vQ(0)\rightarrow 0$).
%  %%%%%%%%%%%%%%%%%%%%% figure
%%%%%%%%%%%%%%%%%%%%%%%%%%%%%%%%%%%%%%%
\begin{figure}

\centerline{\includegraphics[width=.97\columnwidth,angle=0]{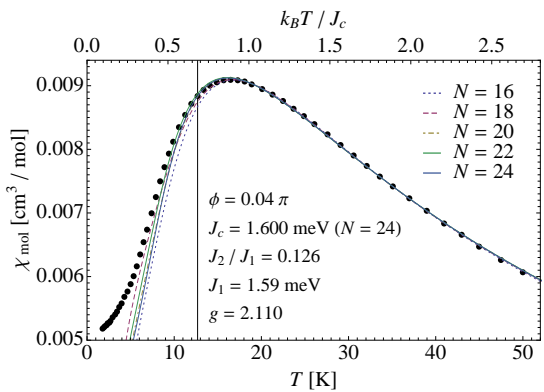}}
\caption{Uniform magnetic susceptibility from FTLM for various tile
sizes. Fitting to experimental
values (dots) for \C{} starts above the vertical line and below the
maximum. Optimal fit parameters $J_1$, $J_2$ 
are indicated. The g-value is obtained from $g\mu_BH_s=4J_1$.}
\label{fig:Fig3}
\end{figure}
%%%%%%%%%%%%%%%%%%%%%%fig%%%%%%%%%%%%%%%%%%%%%%%%%%%%%%%%%%%%%%%%%%

We have checked the results of spin wave theory with an unbiased numerical ED
approach for finite size clusters. The uniform and staggered moment are obtained
from the scaling analysis of the calculated structure factor performed in
analogy to the zero-field case \cite{schmidt:10}. The results are shown in
Fig.~\ref{fig:Fig2} in a parametric representation where \mm{} is plotted versus
\ms . The field is the parameter in the plot starting at H=0 in the lower right
corner (zero field staggered moment) up to $H=H_s$ in the upper left corner
(saturated uniform moment).  The classical reference curve is shown by the
dashed line. The (unscaled) moment values for various sized 2D $J_1-J_2$ tiles
with intermediate frustration angle $\phi/\pi=0.063$ are presented as symbols.
The finite size scaling extrapolation to the thermodynamic limit has to be
performed for constant magnetization given by the horizontal dashed lines. The
result is shown in open and full circles. They agree with the predictions of
spin wave calculations (full lines) for $\phi/\pi=0.04$ and $0.063$. For
comparison the spin wave result for the the nonfrustrated ($\phi=0$) NAF (dotted
line) is also shown. It is suggestive from Figs.~\ref{fig:Fig1}
and~\ref{fig:Fig2} that a careful determination of the ordered moment field
dependence may determine the degree of frustration in a quasi-2D antiferromagnet
given by $\phi$ or $J_2/J_1$.
%  %%%%%%%%%%%%%%%%%%%%% figure %%%%%%%%%%%%%%%%%%%%%%%%%%%%%%%%%%%%%%%%
\begin{figure}

\centerline{\includegraphics[width=.9\columnwidth,angle=0]{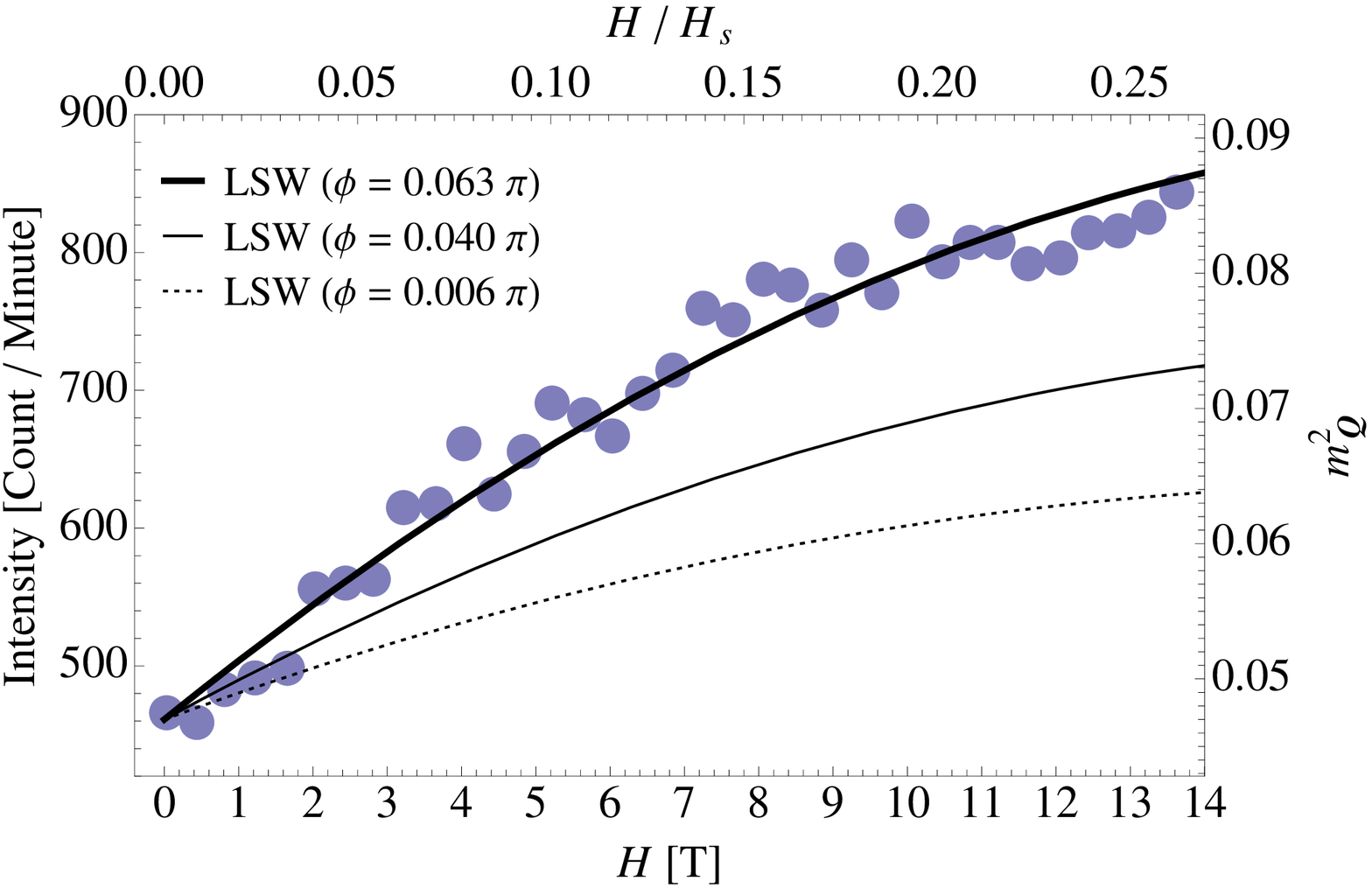}}
\centerline{\includegraphics[width=.9\columnwidth,angle=0]{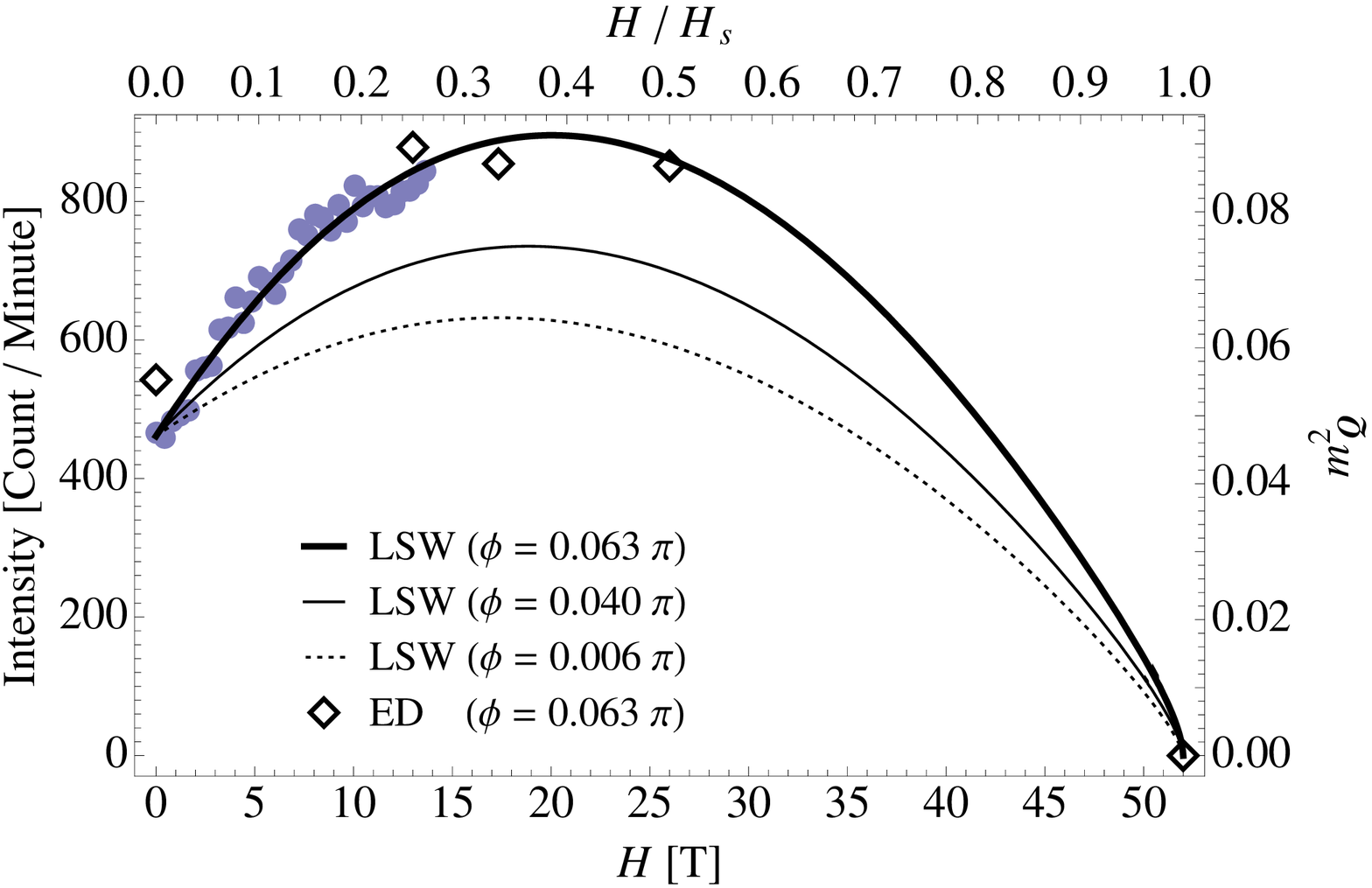}}
\caption{Field dependence of staggered moment squared ($\sim$
scattering intensity) in \C (full circles) (from Ref.~\cite{tsyrulin:10}).
Fat, thin and dotted lines:
Results from linear spin wave theory (LSW) for $\phi/\pi= 0.063, 0.04, 0.006$ respectively
($J_2$/J$_1$= 0.2, 0.126, 0.02). 
Diamonds: ED results for  $\phi/\pi= 0.063$.
Top: Scaled up to the available experimental values. Bottom: Scaled up to the saturation field.}
\label{fig:Fig4}
\end{figure}
%%%%%%%%%%%%%%%%%%%%%%fig%%%%%%%%%%%%%%%%%%%%%%%%%%%%%%%%%%%%%%%%%%%
%

\section{Application to quasi-2D \Cb}
\label{sect:Cupz}

To demonstrate the strength of this method we apply it to the quasi-2D
antiferromagnet \C. In previous work \cite{tsyrulin:09,tsyrulin:10} the spin
wave excitations, field dependent moment and ordering temperature $T_N(H)$ of
this compound were determined. Based on these results  and on earlier
thermodynamic studies \cite{lancaster:07,xiao:09} it was proposed that Cu
pyrazine is close to a pure NAF with a frustration ratio $J_2/J_1=0.02$ or
$\phi/\pi \approx 0.006$. The field dependence of staggered moment however was
not analyzed in this respect. This will be performed in the present work to
illustrate this powerful method for obtaining  $J_2/J_1$. We also compare to the
results of the analysis of $\chi(T)$ using our FTLM data. The previous work
\cite{lancaster:07,xiao:09} based on series expansion assumed from the outset
that $J_2$=0 ($\phi=0$). We perform an unbiased  analysis with possibly non-zero
$J_2$ by fitting the FTLM data with variable $\phi$ and $J_c$ to the
experimental data. In order to reduce the influence of finite size effects only
data points from slightly below the maximum in $\chi(T)$ up to the highest
temperature are included. The result for the best FTLM fit from various cluster
sizes is shown in Fig.~\ref{fig:Fig2} and $J_1$, $J_2$ are given in the caption.
They lead to a best fit with $J_2/J_1 =0.12$ or $\phi/\pi=0.04$. From the value
of $J_1$ and the measured saturation field H$_s=52$ T we get a gyromagnetic
ratio $g=2.11$. For $\phi/\pi>0.04$, the agreement around the maximum of
$\chi(T)$ becomes worse.

However, the fit to $\chi(T)$ does not necessarily yield an accurate and unique
solution, since it depends on $\Theta_{\text{CW}}=J_1+J_2$, but only weakly on
the individual exchange constants~\cite{misguich:03,shannon:04}, and the
complementary values $\phi$ and $\phi'=\frac{\pi}{2}-\phi$ having identical
$\Theta_{\text{CW}}$ corresponding to Néel and columnar AF order cannot be
distinguished from an analysis of $\chi(T)$. It is therefore important to
check this value of $\phi$  by using the field dependence of the ordered moment.
The square of the latter is proportional to the scattering intensity. We
calculated \ms{}  from the spin wave theory  in Eqs.~(\ref{eq:sw})
and~(\ref{eq:ms}) using three different values of $\phi/\pi$. The results are
shown in Fig.~\ref{fig:Fig4}. The staggered moment squared ($m_\vQ^2\sim$ intensity)
increases by about a factor of two in the measured regime up to $H/H_s\simeq
0.25$. Using spin wave calculation (which is accurate for $\phi/\pi\leq 0.1$) supplemented 
by ED  (see also Fig.~\ref{fig:Fig2}) the value
$\phi/\pi=0.063$ $(J_2/J_1=0.2)$ (full line) 
gives perfect agreement with experimental moments.
This is somewhat larger than the value from the FTLM fit to $\chi(T)$.
The experimental data in Fig.~\ref{fig:Fig4} are determined with $7-8 \%$ relative accuracy \cite{tsyrulin:10}.

% For comparison (as in Fig.~\ref{fig:Fig2}) the ED results for the moment are
% shown as open circles. Because the finite size scaling is only possible for
% discrete values of the uniform moment \mm{}  only few values with $h/h_s\leq
% 0.3$ in Fig.~\ref{fig:Fig4} are available from ED.
For comparison we also show the moment for $\phi/\pi =0.04$ (thin line) and the
nearly nonfrustrated NAF with $J_2/J_1 =0.02$ or $\phi/\pi=0.006$ (dotted line).
In the latter model the predicted field induced staggered moment increase is
much too small. To get a more pronounced moment increase with field one has to
increase the frustration $\phi$, as is evident from Fig.~\ref{fig:Fig1}, and for
$\phi/\pi=0.063$ the experimentally observed increase is obtained in
Fig.~\ref{fig:Fig4}. The discrepancy to the FTLM value of $\phi$ may possibly be caused
by the background subtraction process \cite{tsyrulin:10} which gives an
uncertainty to the the absolute size of moment increase.
% In this respect \C{} is not an ideal example due to the relatively large
% non-magnetic background observed. We only mention that analysis of \ms also
% can resolve an ambiguity of $\phi$ possible values. It is known
% \cite{misguich:03,shannon:04} that complementary values $\phi$ and 
% $\phi'=\frac{\pi}{2}-\phi$ corresponding to NAF and CAF phase cannot be
% distinguished from an analysis of $\chi(T)$ or specific heat. However,
% Fig.~\ref{fig:Fig4} shows that \ms for $\phi'/\pi=0.41$ (CAF) (dash-dotted
% line) is quite different from $\phi/\pi=0.04$ (NAF). This observation may be
% used as a tool to resolve the ambiguity in other cases where the ordered
% structure is not known.  %%%%%%%%%%%%%%%%%%%%% figure %%%%%%%%%%%%%%%%%%%%
\begin{figure}

\centerline{\includegraphics[width=.8\columnwidth,angle=0]{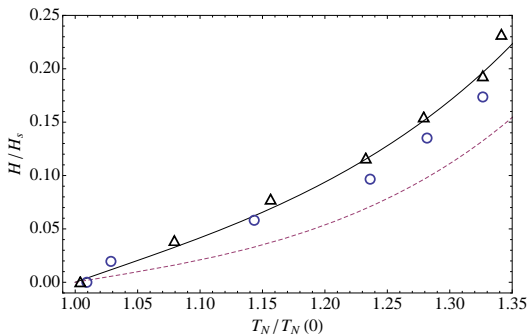}}
\caption{Field dependence of transition temperature from RPA spin
wave theory. Symbols: experimental values \cite{tsyrulin:10}.
Full line: Calculated from Eq.~(\ref{eq:tnh}) with $\phi/\pi=0.063,
J'/J_1 = 6\cdot 10^{-3}$. Dashed line: $\phi/\pi=0.006, J'/J_1 =
6.8\cdot 10^{-4}$.}
\label{fig:Fig5}
\end{figure}
%%%%%%%%%%%%%%%%%%%%%%fig%%%%%%%%%%%%%%%%%%%%%%%%%
%

Quantum fluctuations not only lead to the distinct field dependence
of the ground state staggered moment but are also responsible for the
observed anomalous increase of the N\'eel temperature with field
strength \cite{tsyrulin:10}. The N\'eel order at finite $T_N$ is an
effect of the finite interlayer coupling $J'$. Because of the
quasi-long range order of 2D HAF with exponentially increasing
correlation length a small interlayer coupling $J'/J\ll 1$ leads to
sizable $T_N$ on the scale of the intra-layer exchange strength
\cite{yasuda:05}. $T_N(0)$ for the nonfrustrated model ($\phi =0$) may be obtained from an empirical
formula based on MC simulations, however, this is not available for $\phi\neq 0$ and for finite fields. 
Therefore we use a self-consistent
RPA theory based on quasi-2D spin waves where $T_N(h)$ 
is determined by the condition of vanishing staggered moment. We obtain
\bea
	T_N(h)=\left[
	\frac{4}{N}\sum_\vk
	\left(
	\frac{A_\vk-B_\vk\cos^2\Theta_c}{E_\vk^2(h)}
	\right)
	\right]^{-1}
\label{eq:tnh}
\eea
Now the interlayer coupling along a-direction is included according to
$J_\vk=2J_1(\cos k_y+\cos k_z)+ 4J_2\cos k_y\cos k_z + 2J'\cos k_x$ and 
\vk~ as well as the ordering  vector \vQ~ in $A_\vk, B_\vk$ of Eq.~(\ref{eq:sw})
have now all three components. The integral in Eq.~(\ref{eq:tnh}) is therefore finite leading
to a nonzero $T_N(h)$.
Eq.~(\ref{eq:tnh}) reduces to the expression in Ref.~\onlinecite{majlis:92} for $h=0$. The
RPA theory predicts the right dependence of $T_N(0)$ on $J'/J_1$ but
the absolute values are larger than from those of empirical formulas
fitting the MC simulations \cite{yasuda:05}. Therefore in
Fig.~\ref{fig:Fig5} we have plotted the normalized field  dependent
transition temperature $T_N(h)/T_N(0)$. It is shown for two sets of
values ($\phi,J'$). The increase in $T_N(h)$
is driven by the reduction of quantum fluctuations since in a field
the average spin wave energy is increasing. The $\phi, J'$ values for
the dashed curve reproduce the experimental T$_N(0) = 4.2$ K but
fail for the field dependence. The values derived here (corresponding
to full line)  lead to excellent agreement with experimental
$T_N$(h)/$T_N$(0) but  $T_N(0)$ is about twice the
experimental value.  We think that the functional dependence $T_N(H)$
is more significant for $J'/J_1$ than the single value of $T_N(0)$.

\section{Conclusion}
\label{sect:conclusion}

We have presented the analysis of field dependence of ordered moment
in frustrated quasi-2D antiferromagnets using exact diagonalization
for finite clusters and compared with spin wave theory. The staggered
moment exhibits pronounced nonmonotonic behavior as function of
field which depends on the degree of frustration given by
$\phi=\tan^{-1}(J_2/J_1)$. This provides a
powerful means to extract the frustration ratio which is more
accurate and less ambiguous than using temperature dependence
thermodynamic quantities. We have applied this method to \C{} and
conclude, primarily from the ordered moment field dependence, that it
is a quasi-2D antiferromagnet with intermediate frustration. This method 
may be used more generally for frustrated antiferromagnets. In particular 
it should also be applicable when the field dependent ordered moment is
extracted from analysis  of NMR splittings rather than from neutron diffraction data.
Since it requires the existence of an ordered moment it will, however, not be useful
for compounds corresponding to the disordered regimes of the $J_1-J_2$ phase diagrams, 
if they should indeed exist.

\section*{Acknowledgements}
The authors would like to acknowledge helpful discussion with N. Shannon.

%%%%%%%%%%%%%%%%%%%%%%%%%      References        %%%%%%%%%%%%%%%%%%%%%

%\bibliographystyle{prl}
\bibliography{msh}

\end{document}